\documentstyle[twoside,fleqn,espcrc2]{article}
\def\be{\begin{equation}}
\def\ee{\end{equation}}
\def\bea{\begin{eqnarray}}
\def\eea{\end{eqnarray}}

\newcommand{\AmS}{{\protect\the\textfont2
  A\kern-.1667em\lower.5ex\hbox{M}\kern-.125emS}}

\title{A review of heavy-heavy spectroscopy}

\author{C. T. H. Davies\address{Department of Physics and Astronomy,
        University of Glasgow, \\
        Glasgow G12 8QQ, UK}%
         }

\begin{document}

\begin{abstract}
Calculations of the heavy-heavy spectrum present a good
opportunity for precision tests of QCD using lattice
techniques. All methods make use of a non-relativistic expansion of the
action and its systematic improvement to remove lattice artefacts.
There was convincing demonstration this year that these methods work
and that the associated perturbation theory is well-behaved.
Comparison to experimental results yields an accurate value for
the lattice spacing, $a$, a key result in the determination of $\alpha_s$,
and (for the first time this year) the mass of the $b$ quark (4.7(1) GeV).
\end{abstract}

\maketitle

\section{INTRODUCTION}

The spectrum of heavyonium states is an ideal place to
provide a precision test of QCD from first principles.
Very accurate
calculations are possible using lattice techniques.
The reason is that $both$ statistical and systematic
errors can be
made small, something we are nowhere near achieving
for light hadron calculations.

The key ingredients are
\begin{itemize}
\item A non-relativistic formulation appropriate to the physics.

\item Moderately-sized lattices at moderate values of $\beta$ with
perturbatively improved actions.

\item High statistics from $\cal{O}$(100) configurations and the use of
multi-exponential fits to multiple correlation functions.
\end{itemize}

Results this year demonstrate clearly that perturbatively
improved actions work and give continuum results
when values of the lattice spacing are not very small.
This is particularly useful for heavy quark systems where
lattice discretisation errors are relatively large. For
light hadron systems, however, its use would allow
calculations at lower values of $\beta$ and smaller volumes
than are currently employed, which might be
an advantage. Another clear result is the greatly increased
confidence in the ground state that is obtained from
a multiple exponential fit to multiple correlation functions.
Such a fit is not often possible at the level of
statistics employed in current light hadron calculations.

Results were presented by four groups this year. The groups
differ in the action employed for the heavy quarks, the
level of improvement of the action used, and the gauge field
configurations used. I shall refer to the collaborations
as FNAL~\cite{kronfeld,khadra}, KEK~\cite{onogi},
 NRQCD~\cite{lidsey,sloan}, UKQCD~\cite{drummond,collins}.

The FNAL group~\cite{kronfeld,khadra} use improved heavy Wilson fermions for
both $c\overline{c}$ and $b\overline{b}$ spectra on
quenched gauge field configurations at several values of
$\beta$, but principally using 100 configurations at $\beta$
=6.1.

The KEK~\cite{onogi} group use Wilson fermions for the $c\overline{c}$
spectrum but, for the first time, on gauge field
configurations with 2 flavours of dynamical quarks. They have
75 configurations at $\beta$ = 5.7.

The NRQCD group uses Non-relativistic QCD (NRQCD) to
study $b\overline{b}$~\cite{sloan} (with 100 quenched gauge field
configurations at $\beta=6.0$ from the Staggered
Collaboration) and $c\overline{c}$ spectra~\cite{lidsey} (with 100 quenched
gauge field configurations at $\beta$ = 5.7
from the UKQCD collaboration). The NRQCD action is
improved at next-to-leading order.

The UKQCD collaboration has results for both improved
heavy Wilson~\cite{collins} ($c\overline{c}$ and
extrapolating to $b\overline{b}$) and leading order NRQCD
(for $b\overline{b}$)~\cite{drummond} principally at $\beta$ = 6.2 on
60 quenched configurations.

The most complete spectrum is presented by the NRQCD collaboration,
with both spin-averaged splittings and $s$ and $p$ hyperfine
splittings for $b\overline{b}$ and $c\overline{c}$. I will quote
results from other groups where available. The KEK group
give only the 1S-1P spin-averaged splitting which is useful for
the determination of $a^{-1}$ and $\alpha_s$. Their result provides
a confirmation of the unquenching corrections that other groups
employ to convert their quenched $\alpha_s$ to a physical one. This
is discussed by El-Khadra~\cite{khadra} and I will not refer to the KEK results
further.

\subsection{NRQCD}

NRQCD makes use of an expansion of the quark action in powers
of $v^{2}/c^{2}$ where $v/c$ is a typical velocity
inside the heavy meson. $v^{2}/c^{2}\sim 0.1$ for $\Upsilon$ and
0.3 for $J/\Psi$. The terms in the action can be ordered by
power-counting rules~\cite{nakhleh}.

The lowest order terms are simply
\begin{equation}
S_{lo} = \psi^{\dag}D_t \psi + \psi^{\dag} \frac {D^2} {2M} \psi
\end{equation}
$\psi$ is a 2-component quark field and the separate antiquark has the
same action. The presence of a single time derivative means that
the propagator can be calculated in one sweep through the lattice
using a suitable time evolution equation~\cite{nakhleh} derived
from a lattice version of the action.

This leading order action can already give useful results for
spin-averaged splittings, i.e. splittings in which states of a given
$L$ are averaged over, weighted by the value of $2J+1$. These
splittings can be compared to those extracted from a free
Schr\"odinger equation with the usual central
heavy quark potential. The leading order action has errors of
10\% for $b\overline{b}$ and 30\% for $c\overline{c}$.
The UKQCD collaboration work with this action~\cite{drummond}.

The lowest order spin-dependent terms (those with an explicit
$\sigma$ are
\begin{equation}
S_{\sigma} = \psi^{\dag} \frac {g} {2M} \sigma . B \psi +
\psi^{\dag} \frac {g} {8 M^2} \sigma. (D \times E - E \times D) \psi
\end{equation}
The first of these gives $s$ state hyperfine splittings ($^{3}S_1 - ^{1}
S_0$) and the second $p$ state hyperfine splittings (between $\chi$
states).
Inclusion of these terms gives the hyperfine splittings to 10\%
for $b\overline{b}$ and 30\% for $c\overline{c}$.

The NRQCD group includes also the next-to-leading
contributions to the spin-averaged splittings. These are
given by
\begin{equation}
S_{nlo} = \psi^{\dag} \frac {D^4} {8 M^3} \psi -
\psi^{\dag} \frac {ig} {8M^2} (D.E -E.D) \psi
\end{equation}
In addition there are $\cal{O}$$(a^{2})$ corrections that appear at
the same level and are included through terms
containing $D^{4}$ and $D^{4}_i$~\cite{sloan}.
Now spin-averaged splittings have systematic errors at the 1\% level
for $b\overline{b}$ and at the 10\% level for $c\overline{c}$
from terms neglected in the heavy quark action.

All the terms above have been written with their tree level
coefficients. Potentially large radiative corrections to these
coefficients can be absorbed by a transformation of the
gauge fields :
\begin{equation}
U_{\mu} \rightarrow \frac {U_{\mu}} {u_0}
\end{equation}
This effectively includes in each term tadpole corrections to all
orders~\cite{nakhleh}. These are assumed to be the dominant
corrections~\cite{lepage} so that
any further radiative corrections can be ignored at the next-to-leading
order level. Explicit calculations~\cite{morning} for the coefficients of
$D^{4}$ and $D_i^{4}$ confirm this.

\subsection{Heavy Wilson}

The approach of heavy Wilson fermions is very similar to that of
NRQCD (in fact, eventually identical). It begins by adding
a clover term to the usual Wilson action to correct for
errors at $\cal{O}$$(a)$.
\begin{equation}
S_{clover} = ic \frac {\kappa} {2} \overline{\psi} F_{\mu \nu}
\sigma_{\mu \nu} \psi
\end{equation}
Propagators are calculated by the usual methods for Wilson fermions.
Convergence for heavy quarks (low $\kappa$) is fast but it still
requires $\cal{O}$(10) passes through the lattice. The FNAL group
give $c$ the value calculated from the dominant tadpole effects
described above\cite{kronfeld}. This value is $u_{0}^{3}$ = 1.4 at $\beta$ =
6.1.
The UKQCD group~\cite{collins} give $c$ its tree level value of 1.0.

Compared to the $nlo$ NRQCD approach above, the heavy Wilson
action has terms missing at $\cal{O}$$(a^2)$ and at order
$D^{4}$. However, there is some remnant of the required
$D^{4}$ term present so the error should be less than the
10 \% or 30\% of $lo$ NRQCD.

\subsection{Gauge Field Configurations}

There are two sources of systematic error present in
standard quenched gauge field configurations which must be taken into
account when assessing the results of the FNAL, NRQCD and UKQCD
groups. Luckily the two effects tend to counteract each other to
some extent.

The first is the presence of $\cal{O}$$(a^2)$ errors in the usual
Wilson plaquette gauge action. This can be corrected by the
use of $2\times 1$ of $2\times 2$ plaquettes with an appropriate
tadploe-improved coefficient.

We can estimate the size of these effects from a Schr\"odinger
equation on the lattice using a lattice heavy quark potential
(such as that from Bali and Schilling~\cite{bali}). The $\cal{O}$$(a^2)$ errors
in the gluon propagator show up as a lack of rotational invariance in
the potential at small $R$.
If we correct the $[4sin^2(ka/2)]^{-1}$ of the one-gluon exchange term
to $[4sin^2(ka/2) + 4/3sin^4(ka/2)]^{-1}$ and compare the spectra at fixed
quark mass, we can obtain an estimate of the size of these
$\cal{O}$$(a^{2})$ errors. We find a 2\% reduction of the spin-averaged
$1P-1S$ splitting when the corrections are included.

Another, more serious, source of error is that of the quenching
of the gauge fields (at least partly removed in the KEK results).
We are studying states mainly well below threshold for decay to
heavy-light channels, so the principal effect of quenching
is from the change in the shape of the heavy quark potential at
small distances. The quenched coupling constant runs too fast to
zero, so the potential close to $R=0$ is not deep enough.
Quenching then has the effect of raising states which are concentrated
at short distances, such as $s$ states. A comparison of
results from a lattice Schr\"odinger equation with quenched
and unquenched potentials from the MTc collaboration~\cite{mtc} gave a
shift to the ratio of $2S-1S$ to $1P-1S$ of 10\% (downwards
on unquenching). The potentials did
not have the same lattice spacing so an absolute shift for
$1p-1S$ could not be derived.
We fixed the quark mass in physical units to 4.7 GeV in both cases.

In perturbation theory the $s$ state hyperfine splitting is
proportional to the square of the wavefunction at the origin and
$\alpha_s(M)$. In the above analysis we did not find a big
effect from quenching on the wavefunction at the origin. This disagrees
with earlier results from El-Khadra~\cite{khadra2} who used a continuum
Richardson potential.
We believe that the lattice Schr\"odinger approach may more
accurately reflect what we will find in lattioce simulations.
The effect of quenching on $\alpha_s(M)$ is to reduce it by
20\%. Thus the lattice Schr\"odinger approach predicts an
$s$ state hyperfine splitting 20\% too low in the
quenched approximation and the continuum Richardson gives
a 40\% reduction.

\section{RESULTS}

There are two parameters to be fixed in all these
calculations.

\begin{itemize}
\item $a^{-1}$ : Fix from the spin-averaged spectrum which is relatively
insensitive to $M$. (Compare the $s$ and $p$ splittings for
$b\overline{b}$ and $c\overline{c}$ from the Particle Data Book~\cite{data}).

\item $M$ : Fix from the kinetic mass, $M_2$, of the $\Upsilon$
or $J/\psi/\eta_c$ as appropriate. $M_2$ is given by
the non-relativistic dispersion relation for the energy of a meson
at finite momentum:
\begin{equation}
E(p) = M_1 + \frac {p^2} {2M_2} + \cdots
\end{equation}
\end{itemize}

\subsection{$b\overline{b}$ results}

Figure 1 shows the `spin-averaged' spectrum for $b\overline{b}$
obtained by the NRQCD collaboration at $\beta$ = 6.0 and the UKQCD
collaboration (using leading order NRQCD) at $\beta$ = 6.2.
Since no experimental mass is available for the $\eta_b$ (1S, or
its radial excitations, 2S and 3S), no spin-averaging
of $s$ states is done by NRQCD. The experimental results, given as
bars, are for the $\Upsilon (^{3}S_1)$, 1S, 2S and 3S. The UKQCD
results are inevitably spin-averages since they have no spin-dependent
terms in the action. For the $p$ states a spin-average over the
$^3P_{0,1,2}$ states, all known experimentally at 1P and 2P, is done.
In addition the NRQCD collaboration have a mass for the as yet unseen
$^{1}P_1$ state, the $h_b$.

\begin{figure}[htb]
\vspace{6.0cm}
\caption{The spectrum for $b\overline{b}$ states plotted relative to
the mass of the $\Upsilon$ (1S). Solid horizontal lines mark the
experimental data for the $^3S_1$ and for
the spin-average of the $^{3}P_{0,1,2}$ ($\chi_b$) ground states and radial
excitations. Grey horizontal lines mark the expected masses of
the $^1P_1$ ($h_b$) states. The vertical scale is in GeV.
Squares mark the results from the NRQCD
collaboration. The stars are those from the UKQCD collaboration (using
leading order NRQCD). Their results are all for spin-averaged masses. The
cross on $\Upsilon$(1S) shows that its mass was assumed in the results.}
\end{figure}

By performing a bootstrap fit to the whole spectrum of Figure 1,
the NRQCD collaboration obtain a value for the inverse lattice
spacing, $a^{-1}$ = 2.4(1) GeV. The Q value for the fit is good at 0.4.
Taking the ratio of plaquette values at $\beta$ = 6.0 and
$\beta$ = 6.2 and using lattice perturbation theory gives a
value for the inverse lattice spacing at $\beta$ = 6.2 of
3.2(2) GeV. It is this value of $a^{-1}$ that has been used to
convert the UKQCD results from lattice to physical units.
The FNAL group have calculated so far a value only for the
1P-1S splitting for $b\overline{b}$ and they use this to
determine $a^{-1}$, giving no independent
predictions to appear in Figure 1. They obtain 2.7(2)GeV
 at $\beta$ = 6.1, a value in agreement (using perturbative extrapolations)
with the NRQCD value above.

Notice that all these values of $a^{-1}$ are larger than those
quoted at similar $\beta$ values from the string tension or
light hadron spectra. This simply reflects the scale dependence
of the determination of $a^{-1}$ from the quenched approximation,
when the long distance heavy quark potential is too steep and the
short distance potential not steep enough. The difference is that
the short distance potential can be corrected perturbatively for its
quenching errors.

Figure 1 shows a good fit to experimental data~\cite{data} within the errors.
The statistical and systematic errors of the NRQCD calculation~\cite{sloan}
are such that it is almost possible to see the effect of the quenched
approximation. The 1P level is slightly too low and the 2S slightly
too high. Estimates from a lattice Schr\"odinger equation
described above give a correction for quenching and
gluonic $\cal{O}$$(a^2)$ effects that produces a
very good fit to experiment.

Figure 2 shows the hyperfine spectrum from the NRQCD
collaboration for $b\overline{b}$ 1P
states compared to experiment. Separate results are
given for different components of the $^{3}P_2$ and $^{3}P_1$
states. The fit is good, and provides a stringent test
of the coefficent of the $\sigma.(D \times E)$ term in the
spin-dependent action. Using tree-level coefficients
without correction for tadpoles
would give splittings reduced by a factor of 50\%.
Values for the (unobserved) 1S hyperfine splitting
are 29(4) MeV for NRQCD, 25(4) MeV for FNAL and
18(1) MeV for UKQCD using heavy Wilson with a tree-lvel
improvement coefficient. It seems likely that the UKQCD value
is too low.

\begin{figure}[htb]
\vspace{6.0cm}
\caption{The hyperfine spectrum of the $b\overline{b}$ 1P states,
relative to the spin-average of the $^3P_{0,1,2}$ ($\chi_b$) states.
The horizontal lines mark experimental data; the vertical
scale is in MeV.
The squares are results from the NRQCD collaboration, with error
bars where these would be visible.}
\end{figure}

A pole mass can be extracted for the $b$ quark from a
perturbative analysis of the dispersion relation for the
$\Upsilon$ at finite momentum. A fit to the energy
of propagators at zero and non-zero $p$ gives the
$\Upsilon$ mass in lattice units and its
zero-momentum energy ($M_1$ and $M_2$ in equation 7).
However, $M_1$ and $M_2$ are should be related by
\begin{equation}
M_2 = 2(Z_M M^{0}_b - E_0) + M_1
\end{equation}
where $Z_M$ and $E_0$ are perturbatively
calculable mass renormalisation and energy shift
parameters and $M^{0}_b$ is the bare
heavy quark mass. The NRQCD collaboration~\cite{sloan}
 compares the results for $M_2$
from equation 8 and the value extracted directly using equation
7. They agree to 1\% for a variety of bare heavy quark
masses. This agreement is at the level expected for a
$nlo$ calculation and should be seen as a triumph for the
whole method. It requires both that the $nlo$ coefficients
be correct
and that the lattice perturbation theory is well-behaved. Thus
the assumption of tadpole dominance of the coefficients is correct
and there are no unexpectedly large
perturbative or non-perturbative contributions to $Z_M$ or
$E_0$. The numbers used for $Z_M$ and $E_0$ have been
calculated to $\cal{O}$$(g^2)$ for the action used by
the NRQCD collaboration~\cite{morning} and using the scheme and scale for
$g^2$ advocated by Mackenzie and Lepage~\cite{lepage}.

Fixing $M^{0}_b$ so that the experimental value for
$M_{\Upsilon}$ is obtained then yields a value for
the renormalised or pole $b$ mass, $Z_M M^{0}_b$. This
value is 4.7(1) GeV where the main source of error is
that of higher order corrections to $E_0$. Simulation
errors are not significant.

\subsection{$c\overline{c}$ results}

New results for the $c\overline{c}$ spectrum
were presented only by the NRQCD collaboration. Their results
are shown in Figure 1 of the talk by A.J. Lidsey~\cite{lidsey}. Now
experimental results are available for both the $\eta_c (^1S_0)$ and
the $h_c (^1P_1)$. The spin-averaged spectum is therefore
presented with the spin-averaged $s$ state at zero. There are
two $s$ states below threshold for decay to heavy-light channels.
The 2S state is fit using a 3-exponential fit to 2 correlation
functions. The fit to experiment is good and certainly within
systematic errors of 10\%. An estimate for the position of a $d$
state was obtained by calculating a correlation function for the
$^{1}D_2$. This is compared to the posited $^{3}D_1$ state, $\psi(3770)$.
The $^{1}D_2$ is higher, as expected. There may also be
significant coupling to decay channels here which would tend to
distort the spectrum in the quenched approximation.

Figure 3 shows the $c\overline{c}$ hyperfine spectrum.
Both $s$ and $p$ states are given relative to
their respective spin-averages.
For the $s$ state hyperfine splitting there is some
disagreement between groups. The NRQCD collaboration
obtains a value of 96(4)MeV, the FNAL collaboration, 64(4)MeV and
the UKQCD collaboration, 50(1)MeV. The experimental value is 116MeV.
There is a 30\% systematic error in the calculations from
higher order terms in the quark action and also a systematic
effect from quenching which tends to reduce the value. If the
effect of quenching is a 40\% reduction then the NRQCD value looks
high. If a 20\% effect then the other values look low. The NRQCD result
has the smallest systematic error from the quark action, which
should give the best accuracy in determining the quark mass (something
the hyperfine splitting is sensitive to). On the other hand the FNAL
collaboration have results for the hyperfine splitting for different
values of $\beta$ and see no strong systematic effect from $\cal{O}$$(a^2)$
terms that they have not included. This disagreement needs to be
resolved.

\begin{figure}[htb]
\vspace{6.0cm}
\caption{The hyperfine spectrum of the $c\overline{c}$
 1S and 1P ($\chi_c$) states,
relative to their respective spin-averages.
The horizontal lines mark experimental data; the vertical
scale is in MeV.
The squares are results from the NRQCD collaboration, the circles from
the FNAL collaboration and stars from the UKQCD collaboration (heavy
Wilson). Error
bars are shown where they would be visible.}
\end{figure}

\section{CONCLUSIONS}

Heavy quark spectroscopy can provide a stringent test of QCD.
The $b\overline{b}$ spectrum can be calculated to the 1\% level
using an action correct to $nlo$ in $v^{2}/c^{2}$. Such an
action can be provided within an NRQCD or heavy Wilson approach.
Current results in NRQCD are at this level in the quark action.
There are remaining systematic errors present from the gluon
configurations used.  These should be corrected in future
calculations.

\begin{flushleft}
{\bf Acknowledgements}

I am grateful to all my colleagues in the FNAL, KEK, MTc, NRQCD and UKQCD
collaborations for useful discussions when preparing this talk. I
thank the UK SERC for financial support.
\end{flushleft}

\end{document}